\documentclass[a4paper,11pt]{article}
\pdfoutput=1
\usepackage{jheppub}

\usepackage{graphicx,slashed,xcolor,multirow,bbold,mathtools,sidecap,tikz,bm,ulem,enumitem,booktabs,array}
\usepackage{hyperref,url}

\usepackage[compat=1.1.0]{tikz-feynman}
\usepackage{siunitx,adjustbox}

\definecolor{lime}{HTML}{A6CE39}
\DeclareRobustCommand{\orcidicon}{\hspace{-2.1mm}
\begin{tikzpicture}
\draw[lime,fill=lime] (0,0.0) circle [radius=0.13] node[white] {{\fontfamily{qag}\selectfont \tiny ID}}; \draw[white,fill=white] (-0.0525,0.095) circle [radius=0.007]; 
\end{tikzpicture} \hspace{-3.7mm} }
\foreach \x in {A, ..., Z} {\expandafter\xdef\csname orcid\x\endcsname{\noexpand\href{https://orcid.org/\csname orcidauthor\x\endcsname} {\noexpand\orcidicon}}}

\let\emph\textit

\preprint{PSI-PR-24-15, ZU-TH 34/24}

\title{Explanation of the excesses in associated di-photon production at 152\,GeV in
2HDM}

\author[a,b]{Sumit Banik}
\author[a,b]{Andreas Crivellin\orcidB{}}
\affiliation[a]{Physik-Institut, Universit\"at Z\"urich, Winterthurerstrasse 190, CH--8057 Z\"urich, Switzerland}
\affiliation[b]{Paul Scherrer Institut, CH--5232 Villigen PSI, Switzerland}

\emailAdd{sumit.banik@psi.ch}
\emailAdd{andreas.crivellin@psi.ch}

\abstract{ 
Statistically significant excesses exist at around 152\,GeV in associated di-photon production ($\gamma\gamma+X$) in the sidebands of SM Higgs analyses of ATLAS (using the full run-2 dataset). They are most pronounced in the single-$\tau$, missing-transverse-energy, four-jet and $\geqslant1\ell+\!\geqslant1b$-jet channels ($\approx3\sigma$) and can be explained by the Drell-Yan production of new Higgs bosons, i.e.~$pp\to W^*\to H^\pm H^0$. We first examine the excesses in a simplified model approach, considering that $H^\pm$ decays to $\tau\nu$, $WZ$ or $tb$. Both the $\tau\nu$ and $tb$ decay modes individually lead to a significance of $\lessapprox4\sigma$ while for $WZ$ one can obtain at most $3.5\sigma$. This is because the decays of $WZ$ lead to multiple leptons contributing to the two-lepton channel which does not show an excess at 152\,GeV. Next, we consider two-Higgs-doublet models where the charged Higgs does not decay to $WZ$ at tree-level, finding a significance of $\gtrapprox4\sigma$ for a branching ratio of the new neutral Higgs to photons of $\approx$2\%. Even though this branching fraction is quite sizable, it can be obtained in composite models or via the Lagrangian term $\lambda_6 H_1^\dagger H_1 H_2^\dagger H_1+{\rm h.c.}$ breaking the commonly imposed $Z_2$ symmetry.
}

\newpage
\begin{document}
\maketitle

\section{Introduction} 

Uncovering new (fundamental) particles is one of the primary goals of particle physics. Because the Standard Model (SM) is complete since the Higgs discovery in 2012~\cite{Aad:2012tfa,Chatrchyan:2012ufa,CDF:2012laj}, this would take us into uncharted territory and, hopefully, closer to the ultimate theory describing Nature at the subatomic level. The scalar sector of the SM is particularly promising to contain beyond-the-SM physics because no symmetry or underlying principle enforces its minimality. Furthermore, Higgs bosons naturally couple proportionally to the masses of the SM particles, resulting in LHC constraints which are relatively weak~\cite{ATLAS:2024itc,CMS:2024phk} since valence quarks are nearly massless.

Statistically significant indications for a new Higgs with a mass of 152\,GeV have emerged~\cite{Crivellin:2021ubm,Bhattacharya:2023lmu}\footnote{There are also indirect (i.e.~non-resonant) hints for new Higgses at the electroweak scale, encoded in the LHC multi-lepton anomalies~\cite{vonBuddenbrock:2016rmr,vonBuddenbrock:2017gvy,Buddenbrock:2019tua,Fischer:2021sqw}, including differential $WW$~\cite{Coloretti:2023wng} and top-quark distributions~\cite{ATLAS:2023gsl,Banik:2023vxa}. They are compatible with the 152\,GeV excess and could be related to the 95\,GeV excess~\cite{Coloretti:2023yyq}.}. They are most pronounced in the sidebands of the di-photon search for the SM Higgs in associated production channels ($\gamma\gamma+X$). More specifically, Ref.~\cite{ATLAS:2023omk} studied 22 different signal regions such as missing transverse energy ($X=\rm{MET}$), leptons ($X=\ell$) etc., and Ref.~\cite{ATLAS-CONF-2024-005} contains in addition the single-$\tau$ channel. The most significant excesses at $\approx$152\,GeV are in the $\rm{MET}>100$\,GeV, $\tau$, 4-jet and $\ell b$ (at least one $b$ jet and at least one light lepton) channels with significances of $\approx$3$\sigma$. They are consistent with the Drell-Yan production of an $SU(2)_L$ triplet with $Y=0$~\cite{Ashanujjaman:2024pky}, resulting in a combined significance of $\approx$4$\sigma$~\cite{Crivellin:2024uhc}. 

\begin{figure}[t]
    \centering
    ~~~
    \begin{tikzpicture}[baseline=(current bounding box.center)]
        \begin{feynman}
            \vertex (a);
            \vertex [above left=1.5cm of a] (c) {$q$};
            \vertex [below left=1.5cm of a] (d) {$q$};
            \vertex [right=1.5cm of a] (b) ;
            \vertex [above right=1.5cm of b] (e);
            \vertex [below right=1.5cm of b] (f);      
            \vertex [above right=0.75cm of e] (i) {$\gamma$};
            \vertex [below right=0.75cm of e] (j) {$\gamma$};
            \vertex [above right=0.85cm of f] (k) {$\tau^\pm, W^\pm, \bar b$};
            \vertex [below right=0.85cm of f] (l) {$\nu, Z, t$};
            
            \diagram{
                (d) -- [fermion] (a) -- [fermion] (c);
                (a) -- [boson, edge label=$W^*$] (b);
                (f) -- [scalar, edge label=$H^\pm$] (b) -- [scalar, edge label=$H$] (e);
                (j) -- [boson] (e) -- [boson] (i);
                (l) --  (f) --  (k);     };
        \end{feynman}
    \end{tikzpicture}
    \caption{Feynman diagram showing the Drell-Yan production process $pp\to W^*\to H^\pm H$ with $H\to\gamma\gamma$ and $H^\pm\to tb$, $H^\pm\to \tau\nu$ and $H^\pm\to WZ$.}
    \label{fig:Feynman1}
\end{figure}
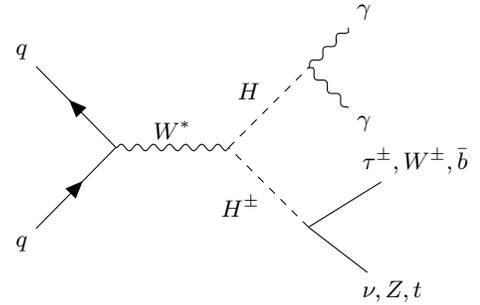

However, while a reasonable agreement was obtained for the preferred signal strength among the different signal regions, i.e.~a $\chi^2$ of $\approx 24$ for $8$ channels, which corresponds to a $\chi^2$ of $3$ per degree of freedom (DoF), for the ``actual" theory describing the data one expects a $\chi^2$ of $1$ per DoF. This motivates searching for a model in which the signal strengths from the different channels agree better, yielding a (even) higher combined significance. Since there is an excess in the single-$\tau$ and single-lepton channels but not in di-$\tau$ and di-lepton signal regions, this suggests that Drell-Yan $pp\to W^*\to HH^\pm$ is the main production mechanism.\footnote{Here $H$ labels any neutral Higgs. In the context of the 2HDM, it could also be the $CP$-odd component $A$.} However, because in the triplet model the relative signal strengths in the single-light-lepton and MET channels are too big compared to the $\ell b$ and $\tau$ channels, one could aim at avoiding the decay of the charged Higgs to $WZ$. This is naturally achieved within a two-Higgs-doublet model (2HDM) where the charged component does not couple at tree-level to gauge bosons.

\begin{figure*}
    \centering
    \includegraphics[scale=0.7]{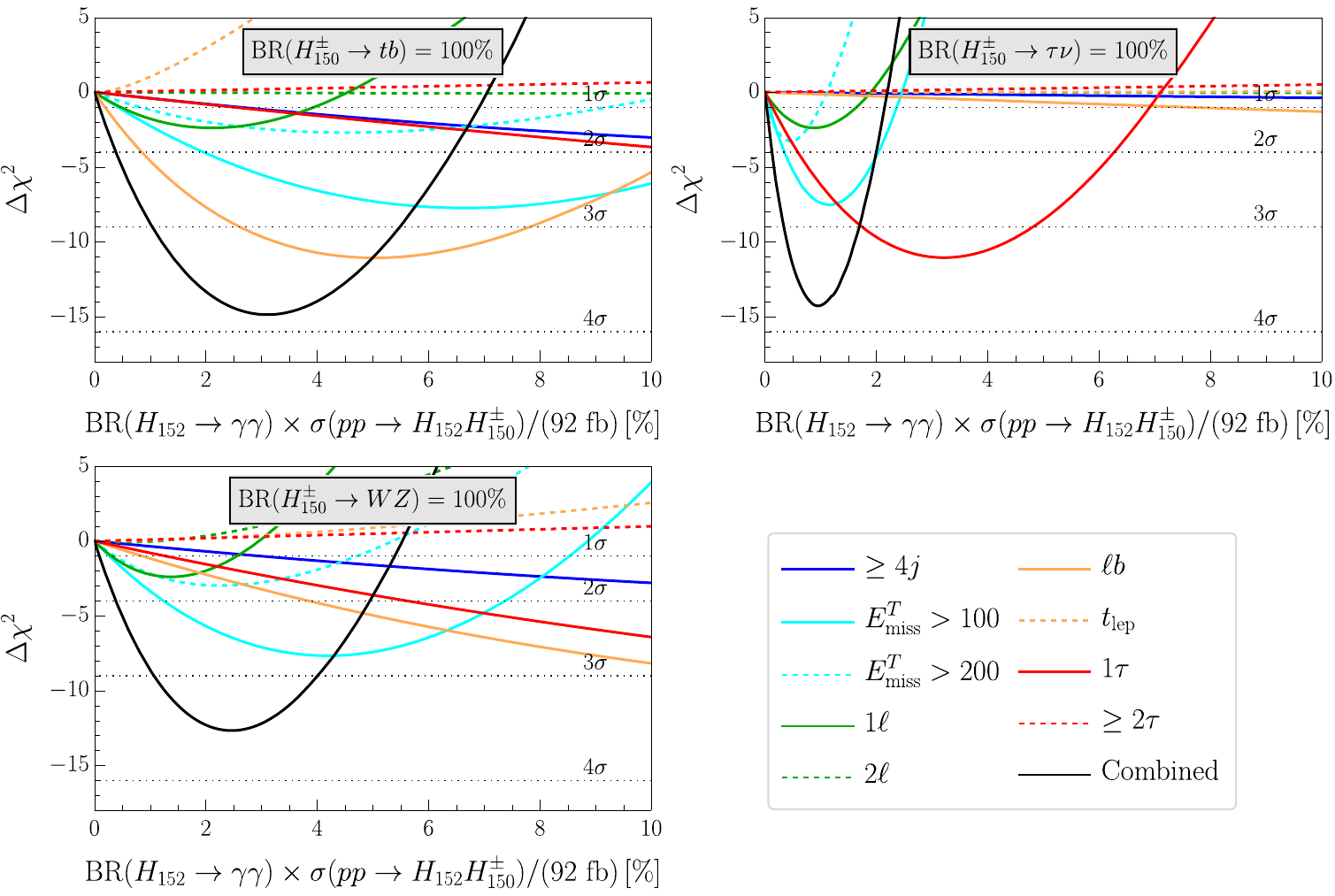}
    \caption{$\Delta \chi^2=\chi^2_{\rm NP}-\chi^2_{\rm SM}$ as a function of BR$(H\to\gamma\gamma)$ for $m_{H^\pm}=150$\,GeV, normalized to the production cross-section of a doublet Higgs of 92fb. Note that the dependence on the charged Higgs mass is quite weak; it mostly leads to a rescaling of the preferred value for BR$(H\to\gamma\gamma)$ due to the varying Drell-Yan production cross-section.}
    \label{fig:chi21d}
\end{figure*}

In the next section, we will perform a simplified model analysis considering the decay channels $H^\pm\to\tau\nu$, $H^\pm\to tb$ and $H^\pm \to WZ$, separately. We will then study the 2HDM where only the first two channels are relevant and perform a combined fit assuming that the charged Higgs couplings to fermions are proportional to the fermion masses (i.e.~type-I like).

\section{Simplified Model Analysis}

As motivated in the introduction, we consider the Drell-Yan production $pp\to W^*\to HH^\pm$ with $m_H=152$\,GeV (see Fig.~\ref{fig:Feynman1}). This is the most important production mechanism if the new Higgs transforms non-trivially under $SU(2)_L$, has a small vacuum expectation value, no direct (or tiny) Yukawa couplings and small mixing with the SM Higgs. Furthermore, one can search for this process in associated di-photon production ($\gamma\gamma+X$). This additional requirement significantly reduces the SM background and enhances the new physics sensitivity.

The dominant decay modes of Higgs boson are generally expected to be the heaviest SM particles, albeit suppression by kinematic factors. More specifically, for the charged Higgs we have the decay modes $H^\pm\to tb$ (including that the top quark can be off-shell) and $H^\pm \to\tau\nu$ (for electroweak scale masses) if it originates from an $SU(2)_L$ doublet as well as $WZ$ for the triplet case. For these cases the relevant di-photon associated production channels ($X$) are light leptons ($X=1\ell$ and $\geq2\ell$), taus ($\tau$), leptonic top decays ($t_{\rm lep}=1\ell +1b$), $\ell b$ ($\geq1\ell +\geq1b$), missing energy (${\rm MET}>100$ GeV and ${\rm MET}>200$ GeV) as well as four jets.\footnote{The signal region targeting hadronic top decays is not included because ATLAS uses a boosted decision tree (BDT) which targets top-pair production via a tight cut on the BDT score of 0.9. Our signal with a bottom quark and an off-shell top quark is quite different resulting in a very small efficiency. We also used the single lepton category from Ref.~\cite{ATLAS-CONF-2024-005}, rather than from Ref.~\cite{ATLAS:2023omk}, since the bottom-quark jet veto leads to a nearly uncorrelated data set w.r.t.~the $\ell b$ category of Ref.~\cite{ATLAS:2023omk}.} They are shown in Fig.~\ref{fig:fitted_eventdistribution} in the appendix. 

We simulated the processes $pp\!\to\! W^*\to (H^\pm\!\to\! \tau\nu,tb,WZ)(H\!\to\! \gamma\gamma)$ using {\tt MadGraph5aMC@NLO}~\cite{Alwall:2014hca}. The parton showering was performed by {\tt Pythia8.3}~\cite{Sjostrand:2014zea} and the ATLAS detector~\cite{ATLAS:2023omk} was modeled with {\tt Delphes}~\cite{deFavereau:2013fsa}. The UFO model file at next-to-leading order (NLO)~\cite{wu_yongcheng_2023_8207058} was built using {\tt FeynRules}~\cite{Degrande:2014vpa}. We normalize the production cross section $\sigma (pp\to W^*\to HH^\pm)$ to the one of an $SU(2)_L$ doublet, with $m_{H^\pm}=150$\,GeV, of 92\,fb, obtained by rescaling the cross section calculated with {\tt MadGraph5aMC@NLO} by the NNLL and NLO QCD correction factor of Refs.~\cite{Ruiz:2015zca,AH:2023hft}. 

\begin{figure*}
    \centering
    \includegraphics[scale=1.15]{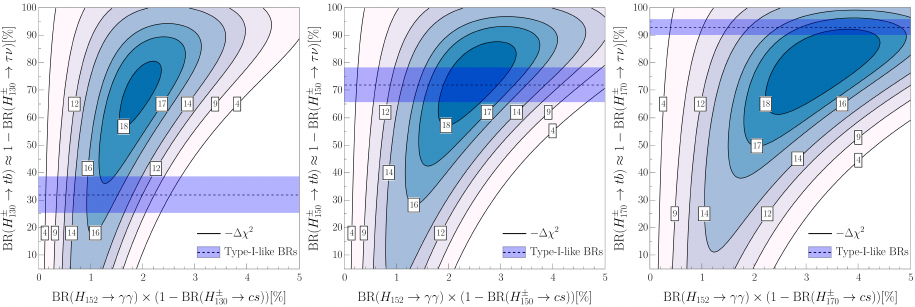}
    \caption{Contours of $-\Delta \chi^2=\chi^2_{\rm NP}-\chi^2_{\rm SM}$ for $m_H=152$\,GeV and $m_{H^\pm}=130$\,GeV, 150\,GeV and 170\,GeV assuming that $\tau\nu$ and $tb$ are the dominant decay modes of the charged Higgs. Because the decay to $cs$, which can be numerically sizable, has a very small impact on all signal regions, its effect can be reabsorbed into a rescaling of BR$(H\to \gamma\gamma)$.  }
    \label{fig:taunutb}
\end{figure*}

For fitting the signal and extracting the significance for the 152\,GeV Higgs, we follow the analysis of Ref.~\cite{Crivellin:2024uhc}. We subtract from the measured ATLAS data the prediction from the SM Higgs signal events as well as the events due to the 152\,GeV Higgs and describe the remaining continuous background by the function
\begin{equation}
    \left(1 - \frac{m}{\sqrt{s}}\right)^{b} + (m/\sqrt{s})^{a_0 +  a_1\log (m/s)}\,.\label{background}
\end{equation}
The free parameters $a_0, a_1$ and $b$ are fitted, $\sqrt{s}=13$\,TeV is the LHC run-2 center-of-mass energy, and $m$ the invariant mass of the di-photon pair. We then perform a simultaneous log-likelihood ratio ($\mathcal{L}_R$) fit using Poisson statistics and convert it to a $\chi^2$ function via $\Delta\chi^2=-2\ln(\mathcal{L}_R)$.

Starting with the assumption BR$(H^\pm \to tb)=100 \%$, one can see from Fig.~\ref{fig:chi21d} that the dominant associated production channels are $\ell b$, $t_{\rm lep}$, $1\ell$ and MET. Since there is no indication for a signal in $t_{\rm lep}$ at 152\,GeV, the best-fit value BR($H\to\gamma\gamma$)\,$\approx$\,3\% leads to a significance of around $3.8\sigma$. Also, note that $H^\pm \to tb$ only gives a small effect in the $\tau$ channel and thus cannot explain the corresponding excess.

For the case of BR$(H^\pm \to \tau\nu)=100 \%$, the $1\ell$ and MET channels, receiving contributions from leptonic tau decays, prefer a smaller BR to photons than the single-$\tau$ channel. This means that the $1\ell$ and MET channels must display an under-fluctuation and/or that the $\tau$ channel contains an over-fluctuation. However, the tensions are within the expected range for statistical fluctuations and are included in the combined significance of around $3.8\sigma$ at BR($H\to\gamma\gamma$)$\approx$1\%. As expected, the decay mode $H^\pm \to \tau\nu$ cannot address the excesses in the $\ell b$ and 4$j$ signal regions.

Finally, the $WZ$ channel results in dominant effects in $1\ell$, $2\ell$ (where no excess is seen) and MET channels with a combined significance of $3.5\sigma$ at BR($H\to\gamma\gamma$)$\approx$2.5\%. However, it has only a small contribution in the single-$\tau$, 4$j$ and $\ell b$ channels.

Let us now look at possible combinations of the charged Higgs decay channels. As we already studied the case with all three charged Higgs decay channels in Ref.~\cite{Crivellin:2024uhc} for the triplet with $Y=0$, finding a significance of $\approx$4$\sigma$, let us therefore focus on combining only $H^\pm\to \tau\nu$ and $H^\pm\to t^*b$. They are complementary in their effects in the $\ell b$ and $\tau$ signal regions and the dominant modes in 2HDMs with natural flavour conservation~\cite{Glashow:1976nt,Paschos:1976ay}, MFV~\cite{DAmbrosio:2002vsn,Botella:2009pq} or alignment~\cite{Pich:2009sp,Tuzon:2010vt}. Therefore, we consider the BR$(H^\pm\to tb)$--BR$(H^0\to \gamma\gamma)$ plane in Fig.~\ref{fig:taunutb}. This means we assume that $tb$ and $\tau\nu$ are the dominant decay modes of the charged Higgs such that BR$(H^\pm\to tb)$+BR$(H^\pm\to \tau \nu)\approx1$. Note that $H^\pm\to cs$ does not lead to a relevant effect in any signal region, such that its presence can be included by rescaling BR$(H^0\to \gamma\gamma)$. For $m_{H^\pm}=130\,$GeV, $m_{H^\pm}=150\,$GeV and $m_{H^\pm}=170\,$GeV, one can see that with an increasing mass, the $-\Delta \chi^2$ maximum slightly increases. The reason for this is the relative efficiency in the $\ell b$ channel (where an excess is seen) compared to the $t_{\rm lep}$ channel (where no indication for an excess at 152\,GeV is seen) is enhanced. Furthermore, with an increasing charged Higgs mass a higher branching ratio to photons is preferred, mainly to compensate for the decreasing Drell-Yan production cross-section. The blue horizontal bands, whose widths are due to the sizable uncertainty of the decay width to off-shell top quarks~\cite{LHCHiggsCrossSectionWorkingGroup:2016ypw}, show the one-dimensional case in which the charged Higgs couplings are proportional to the masses of the involved fermions, as in the case of the type-I 2HDM. In this setup, a combined significance of $>4\sigma$ can be obtained for $m_{H^\pm}=150\,$GeV.
 
\section{\label{sec:model} 2HDM Phenomenology}

\begin{figure*}
    \centering
    \includegraphics[scale=0.55]{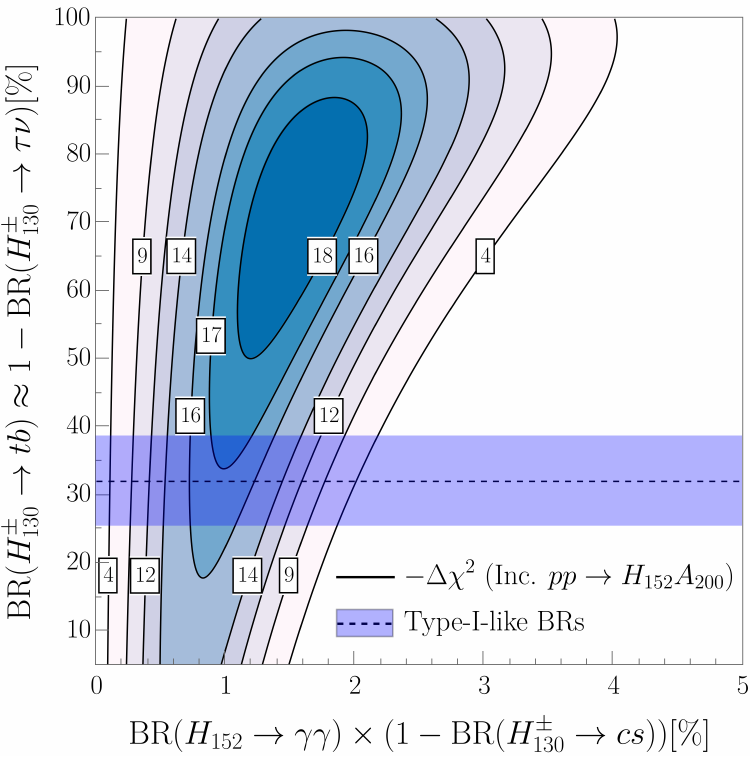}
        \includegraphics[scale=0.55]{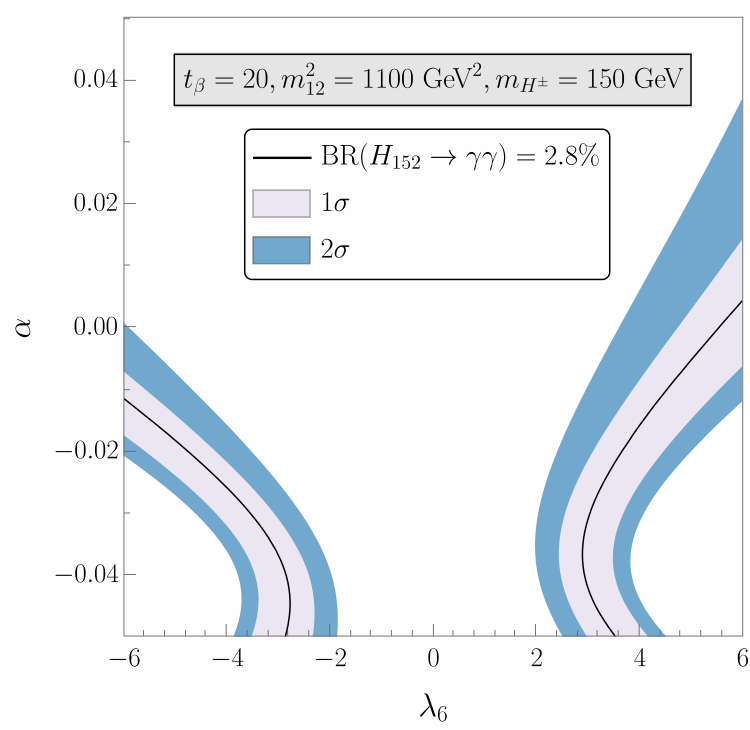}
    \caption{Left: Same as Fig.~\ref{fig:taunutb} but including $pp\to Z^*\to AH$ for $m_A=200$\,GeV and BR($A\to WH^\pm\approx100$\%). Right: Preferred $1\sigma$ and $2\sigma$ regions in the $\alpha$--$\lambda_6$ plane, assuming $\lambda_6$ to be real for $\tan(\beta)=20$. The minimal $\Delta\chi^2$ is $\approx -23$ corresponding to $4.4\sigma$ for this two-dimensional hypothesis.}
    \label{fig:ppZAH}
\end{figure*}

So far, we have studied a simplified model, considering $pp\to W^*\to HH^\pm$ with BR$(H\to\gamma\gamma)$ being a free parameter. Let us now examine the excesses at 152\,GeV within 2HDMs~\cite{Lee:1973iz,Haber:1984rc} where, after spontaneous symmetry breaking, an additional $CP$-odd Higgs boson ($A$) emerges (for more details and our conventions see e.g.~Ref.~\cite{Branco:2011iw}). To study the effect of $pp\to Z^*\to HA$, which we disregarded so far, we consider the benchmark point $m_{H^\pm}=130\,$GeV and $m_A=200$\,GeV with small Yukawa couplings of $A$ to fermions such that BR$(A\to W H^\pm)\approx100$\%. The resulting impact can be seen by comparing Fig.~\ref{fig:ppZAH} (left) to Fig.~\ref{fig:taunutb} (left). The $-\Delta \chi^2$ slightly increases since $pp\to Z^*\to HA$ with $A\to W (H^\pm\to tb)$ has a higher relative efficiency in $\ell b$ vs $t_{\rm lep}$ signal region than $pp\to Z^*\to H(H^\pm\to tb)$.

A branching ratio of $H$ to photons at the percent level is needed to explain the excesses. This cannot be achieved in plain $Z_2$-symmetric 2HDM with natural flavour conservation, since decays to gauge bosons and fermions are correlated to photon decays, because it is induced by loops involving them. To circumvent this, one could add the effective operator
\begin{equation}
F^{\mu\nu}F_{\mu\nu} \phi_1^\dagger \phi_2. + \text{h.c.}\,,
\end{equation}
which dominantly contributes to the decays of the new Higgs bosons for $\tan\beta=v_2/v_1\gg 1$. This corresponding Wilson coefficient is generated in composite Higgs models which can naturally lead to large decay widths of the new Higgses to photon~\cite{Cacciapaglia:2021agf,Benbrik:2019zdp,Cacciapaglia:2019bqz,Belyaev:2016ftv} due to the new (excited) states in the loop. In this case, fermion loops are in general dominant and one predicts $\Gamma(A\to\gamma\gamma)=\frac{9}{4}\Gamma(H\to\gamma\gamma)$
for $m_A=m_H$.

Alternatively, one can give up on the usually employed $Z_2$ symmetry and include in the Lagrangian
\begin{equation}
\mathcal{L} \in -\lambda_6 H_1^\dagger H_1 H_2^\dagger H_1+{\rm h.c.}\,,.
\end{equation}
This term leads to a modification of the width of $H\to \gamma\gamma$ via the charged Higgs loop with the amplitude
\begin{align}
   & \mathcal{M}(H\to\gamma\gamma) \propto \frac{\alpha_{\rm em} g_2 v^2 \lambda_6}{2 m^2_{H^\pm}} \beta_H^0\left(\frac{4 m^2_{H^{\pm}}}{m^2_H}\right),\\ \nonumber &   \beta^0_H(x) = -x\left[1-xf(x)\right],\; f(x) = (\sin^{-1} (1/\sqrt{x}))^2 
\end{align}
for $x \geq 1$~\cite{Spira:2016ztx}, in the limit of $\tan\beta\to\infty$, but not to the SM Higgs. 

We illustrate this in Fig.~\ref{fig:ppZAH} (right) where the preferred regions in the $\alpha$--$\lambda_6$ plane for $\tan\beta=20$ and $m_{12}^2=1100 \, {\rm GeV}^2$ in the type-I 2HDM are shown.\footnote{Note that without the commonly used $Z_2$ symmetry one cannot impose natural flavour conservation, but flavour-changing neutral currents can be avoided via MFV~\cite{DAmbrosio:2002vsn,Botella:2009pq} or alignment~\cite{Pich:2009sp,Tuzon:2010vt}. Furthermore, in the aligned 2HDM one can effectively recover the type-I Yukawa structure in the limit of large couplings $\xi_f$ (see Refs.~\cite{Pich:2009sp,Tuzon:2010vt}). We checked with ScannerS~\cite{Coimbra:2013qq,Muhlleitner:2020wwk}, based on HDECAY~\cite{Djouadi:1997yw,Djouadi:2018xqq} and EVADE~\cite{Hollik:2018wrr}, that vacuum stability and perturbativity that $A\to WH$ is, in fact, the dominant decay mode for our benchmark point.} One can see that it is possible to obtain the preferred size of the branching ratio to photons for order-one values of $\lambda_6$\footnote{{We checked using the Mathematica code of Ref.~\cite{Kannike:2016fmd} that for our benchmark point with order-one values of $\lambda_6$ is consistent with vacuum stability constraints as long as $\lambda_7$ is positive.}}. The resulting $\Delta\chi^2$ minimum is $\approx-23$ (corresponding to 4.4$\sigma$ for our 2-dimensional hypothesis), including the contribution to $\chi^2$ from the constraints of the inclusive $\gamma\gamma$ channel~\cite{Crivellin:2021ubm}. 

\section{Conclusions and Outlook}

We study possible simplified-model explanations of the excesses at 152\,GeV in associated di-photon production via the Drell-Yan process $pp\to W^*\to HH^\pm$ for the three decay modes $H^\pm \to \tau\nu$, $H^\pm \to tb$ and $H^\pm \to WZ$. The $H^\pm \to \tau\nu$ decay channel predicts interesting correlations between the single-$\tau$, single-lepton and MET signal regions. Furthermore, its signatures in the signal regions which show excesses, are complementary to the ones of the $H^\pm \to tb$ decay, which can address the $\ell b$ channel. On the other hand, the $H^\pm \to WZ$ decay in general leads to too many leptons because no excess in the two-lepton signal region is observed.

Therefore, we considered in a second step 2HDMs where the decay $H^\pm \to WZ$ is not allowed at tree-level. We find that for $m_{H^\pm}\approx 150$\,GeV, assuming it has type-I-like BRs, one can obtain a combined significance of $4.3\sigma$, which even slightly increases if $pp\to Z^*\to HA$ with $A\to WH^\pm$ is included. Because a large BR($H\to\gamma\gamma$) at the percent level is required, this suggests supplementing the 2HDM with higher dimensional operators which could be realized within a composite setup. Alternatively, the Lagrangian term $\lambda_6 H_1^\dagger H_1 H_2^\dagger H_1$ can lead to sizable BR in the alignment limit ($\alpha\to0$, $\beta\to \pi/2$) and, including the inclusive di-photon channel, result in a significance of $4.4\sigma$. While we considered $m_H=152$\,GeV, also $A$ could be the 152\,GeV boson. In this case, an imaginary part of $\lambda_6$ would be needed to modify BR$(A\to\gamma\gamma)$, giving at the same time rise to interesting correlated effects in electric dipole moments.

\begin{acknowledgments}
We thank Sezen Sekmen for useful discussions and Saiyad Ashanujjaman, Guglielmo Coloretti, Siddharth P.~Maharathy and Bruce Mellado for agreeing to reuse some results of Ref.~\cite{Crivellin:2024uhc}. This work is supported by a professorship grant from the Swiss National Science Foundation (No.\ PP00P21\_76884).
\end{acknowledgments}

\appendix

\section{Fit to the individual signal regions}

\begin{figure*}[htb!]
\centering
\includegraphics[scale = 0.4]{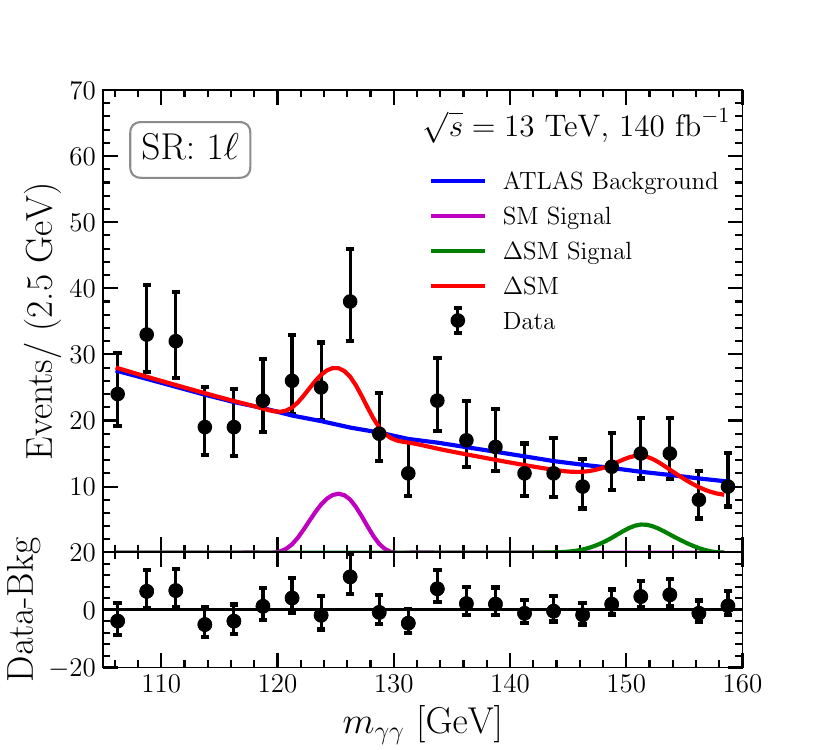}
\includegraphics[scale = 0.4]{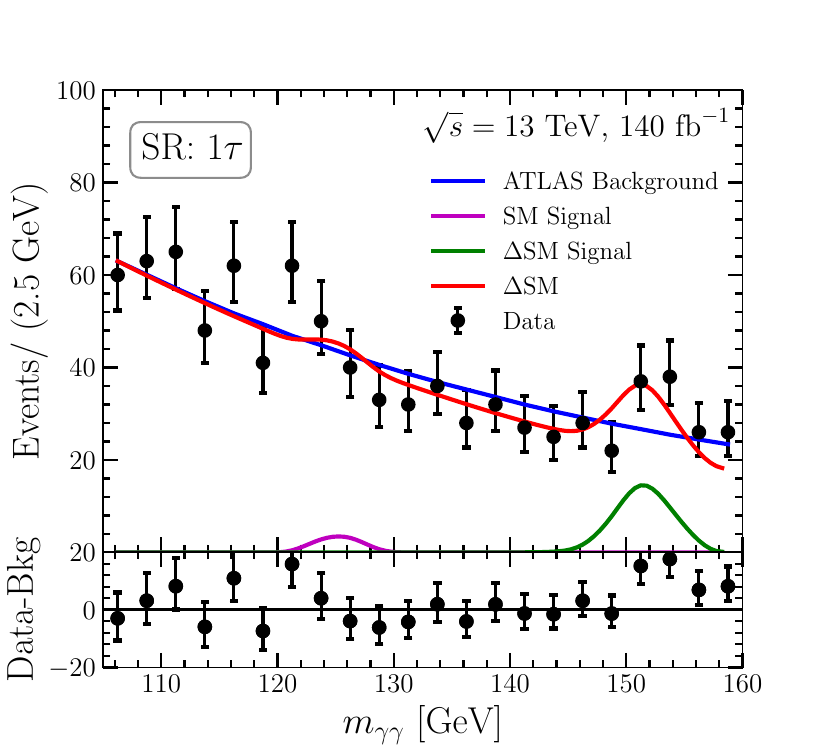}
\includegraphics[scale = 0.4]{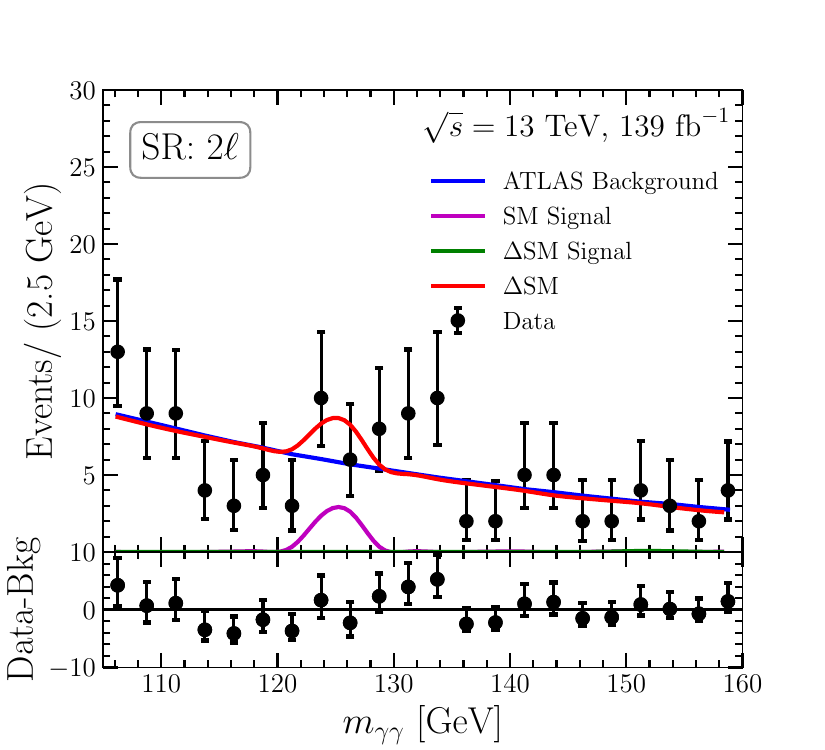}
\includegraphics[scale = 0.4]{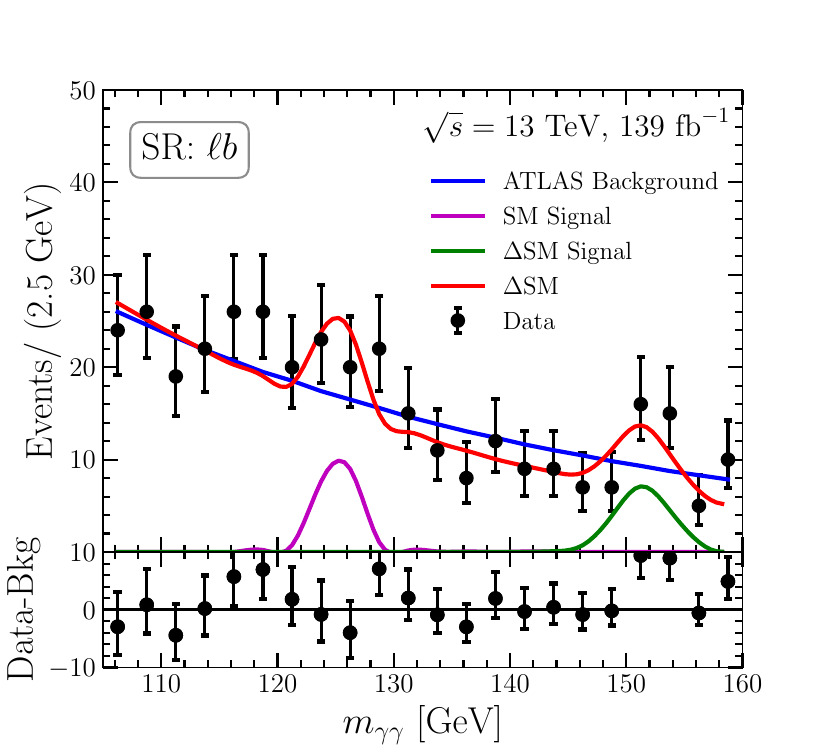}
\includegraphics[scale = 0.4]{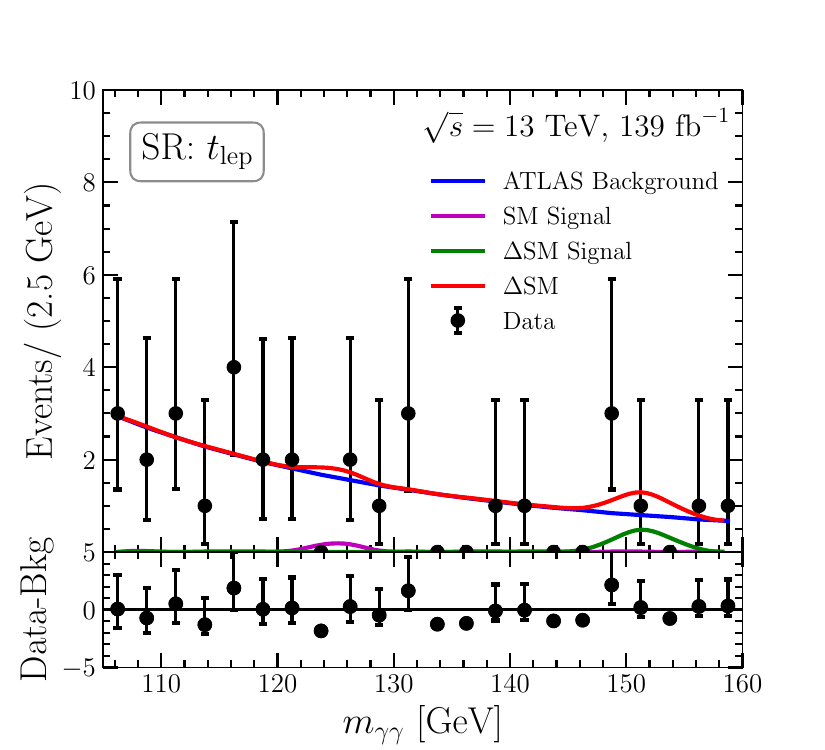}
\includegraphics[scale = 0.4]{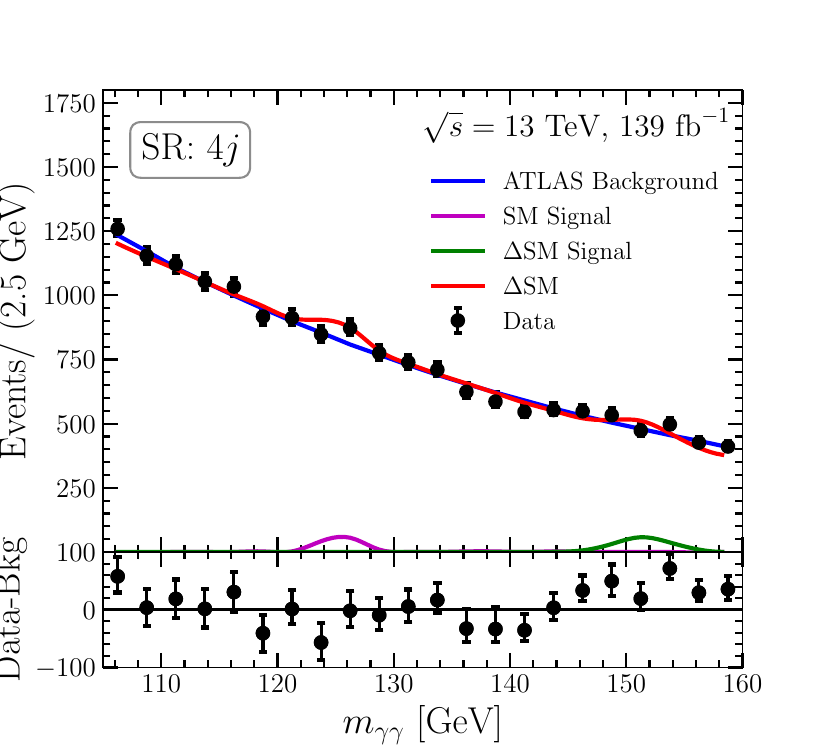}
\includegraphics[scale = 0.4]{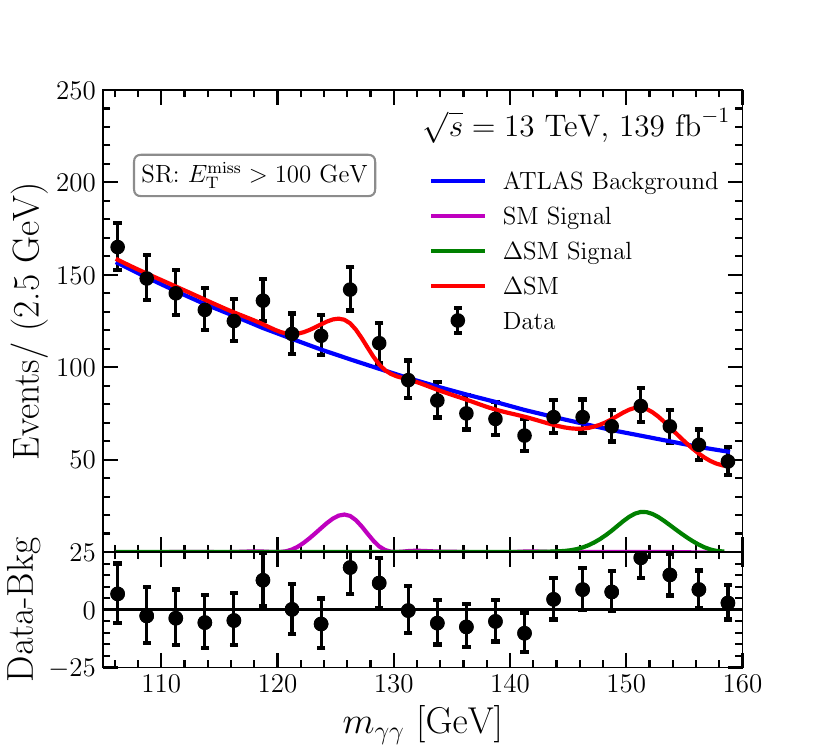}
\includegraphics[scale = 0.4]{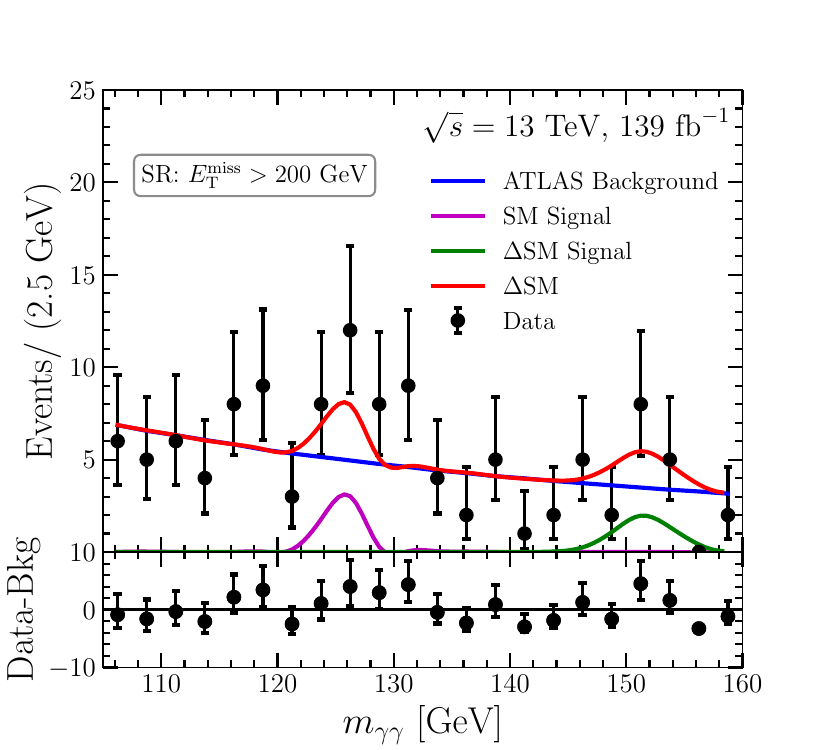}
\caption{Di-photon invariant mass distributions for eight relevant signal regions from Ref.~\cite{Crivellin:2024uhc}. The data (black) is shown together with the continuum background (blue) given in the ATLAS analyses and the total number events (red). The latter contains the refitted background (not shown for brevity), the predicted SM Higgs signals at 125\,GeV (magenta) and the individual best fits for a 152\,GeV signal (green).}
\label{fig:fitted_eventdistribution}
\end{figure*}

We show the eight relevant signal regions of the ATLAS analyses~\cite{ATLAS:2023omk,ATLAS-CONF-2024-005} in Fig.~\ref{fig:fitted_eventdistribution} with the individual best fits for a 152\,GeV Higgs signals~\cite{Crivellin:2024uhc}.

\bibliographystyle{utphys}

\bibliography{apssamp}

\end{document}